\documentclass[preprint]{aastex}

\shorttitle{Collisional Equilibrium}
\shortauthors{O'Brien and Greenberg}

\newcommand{\diffd}{\mathrm{d}}

\usepackage{amsmath}
\usepackage{natbib}

\begin{document}

\title{Steady-State Size Distributions for Collisional Populations:  Analytical Solution with Size-Dependent Strength}

\author{David P. O'Brien and Richard Greenberg}
\affil{Lunar and Planetary Laboratory \\ University of Arizona \\ Tucson, AZ 85721-0092}

\affil{\vspace{0.2in} \it Published in Icarus, August 2003 (Volume 164, Issue 2, pp.~334-345)}
\affil{\it Figures at end of Paper}

\section*{ABSTRACT}

The steady-state population of bodies resulting from a collisional cascade depends on how material strength varies with size. We find a simple expression for the power-law index of the population, given a power law that describes how material strength varies with size.  This result is extended to the case relevant for the asteroid belt and Kuiper belt, in which the material strength is described by 2 separate power laws---one for small bodies and one for larger bodies.  We find that the power-law index of the small body population is unaffected by the strength law for the large bodies, and vice versa.  Simple analytical expressions describe a wave that is superimposed on the large body population because of the transition between the two power laws describing the strength.  These analytical results yield excellent agreement with a numerical simulation of collisional evolution.  These results will help to interpret observations of the asteroids and KBOs, and constrain the strength properties of those objects.  

\noindent \textit{Keywords:} Asteroids; Collisional Evolution.

\section{Introduction}

To interpret the statistics of main-belt asteroids, \citet{1969JGR....74.2531D} analytically modeled a population of self-similar bodies (same collisional response parameters, such as strength per unit mass) in a collisional cascade and found that the steady-state power-law index of the differential size distribution of such a population is $3.5$ (actually $-3.5$, but the absolute value is more commonly quoted).  That model included debris from both cratering and catastrophic shattering events, but concluded that the effect of cratering debris is negligible.  \citet{1989aste.conf..778G} constructed a simple analytical model that includes only catastrophic fragmentation which also yields a steady-state population index of $3.5$.  The value of the population index, $3.5$, is independent of many of the parameters describing the fragmentation process, such as the power-law index of the fragment distribution in a catastrophic collision.  Even if the fragment distribution varies with impact energy, as is seen in laboratory and numerical experiments (i.e.~more energetic collisions on a given body give a steeper fragment distribution), the steady-state population index remains $3.5$ \citep{1994Icar..107..117W}. \citet{1996Icar..123..450T} showed this to be true for any fragmentation model that is independent of the size of the target.  Thus, a value of $3.5$ is frequently cited as the expected steady-state power-law index of a collisionally evolved population, such as the asteroid belt.

This value of the population index, however, is based on the assumption that bodies of every size have the same strength per unit mass.  Analytical scaling arguments \citep{1985Icar...63...30D, 1982Icar...52..409F, 1990Icar...84..226H, 1994P&SS...42.1067H} as well as numerical modeling \citep{1992PhDT.........4R, 1996Icar..124..141L, 1997Icar..129..562M, 1999Icar..142....5B} and laboratory studies \citep{1999Icar..142...21H} have shown that material strength is in fact a size-dependent property.  For bodies smaller than $\sim 1$ km, material properties cause strength to decrease with increasing size.  For larger bodies, strength increases with size due to gravitational self-compression and the gravitational reaccumulation of collisional fragments.  

Numerical collisional evolution models have found that in general, the power-law index of the population is larger than $3.5$ when strength decreases with increasing size and smaller than $3.5$ when strength increases with increasing size, hence the small body population where material properties dominate the strength should be steeper than the large body population where gravity dominates \citep{durda:thesis,1994P&SS...42..599D, 1997Icar..130..140D, 1998Icar..135..431D}.  In addition, all of these researchers found that the transition between these two different regimes creates waves which propagate through the large body size distribution.  

Here we derive those results analytically.  In Section 2, we derive an expression for the steady-state power-law index of a collisional population in which the material strength varies with size as a single power law and show that the canonical $3.5$ value only holds for constant strength.  In Section 3, we show that if the strength is described by a jointed power law (i.e.~decreasing with size for small bodies and increasing with size for large bodies), the population indices in these two regimes are the same as would be calculated in the single-slope case---this result implies that, in terms of the population index, the large body population has no effect on the small body population and vice versa.  We also confirm analytically that for the jointed power law case, waves are introduced in the large end of the population as perturbations about a power law, and we derive simple expressions for their amplitude and wavelength.

In Section 4, we compare the results of our analytical model with a numerical collisional evolution model and obtain excellent agreement.  Finally, in Section 5 we use our analytical results to determine the strength law needed to fit the asteroid belt, and compare our results to results in the literature from collisional modeling, analytical theory, and experiments.

\section{Single-Slope Collisional Model}

First consider the steady-state of a colliding population of bodies whose strength is described by a single power law.  The population is described by the power law:

\begin{equation}
	\label{popeq}
	\diffd N = B D^{-p} \diffd D 
\end{equation}

\noindent where $\diffd N$ is the incremental number of bodies in the interval $[D, D+\diffd D]$.  As there are more small bodies than large bodies, the coefficient $B$ should technically be negative.  However, in the context of an incremental size distribution, $B$ is defined to be positive to avoid physically unrealistic notion of negative numbers of bodies in a given size interval.  $p$ is the power-law index of the population, or simply, the `population index.'  On a log-log plot, Eq.~\ref{popeq} would plot as a line with a slope of $-p$.  For the \citet{1969JGR....74.2531D} solution, $p=3.5$.

\subsection{Collisional Destruction}

The criterion for an impact to result in catastrophic disruption is conventionally parameterized by the the critical specific energy $Q^*_D$, which is defined as the minimum collisional energy per unit mass of the target to fragment the target and disperse half of its mass to infinity (this standard notation was adopted at the 5th Workshop on Catastrophic Disruption in the Solar System, 1998).  By definition, an impact occurring with a specific energy of exactly $Q^*_D$ will yield a largest remaining fragment that is a fraction 
 $f_l=0.5$ of the target mass.  Impacts occurring with a specific energy greater than $Q^*_D$ will give a smaller collisional remnant ($f_l<0.5$).  We assume that impacts occurring with a specific energy less than $Q^*_D$ contribute relatively few collisional fragments, consistent with laboratory work indicating that there is a relatively abrupt transition between small crater formation and widespread target damage \citep{1977Icar...31..277F}.  As shown by \citet{1969JGR....74.2531D}, cratering debris have a negligible effect on the steady-state size distribution.
	      
Here we are concerned with defining the rate at which bodies of diameter $D$ are disrupted (in Section 2.2 we parameterize the size distribution of small bodies produced by such an event).  The diameter $D_{dis}$ of the smallest body that can catastrophically disrupt a target of diameter $D$ can be found by equating the kinetic energy of the projectile to the total energy required for disruption (assuming the same density of projectile and target):

\begin{equation}
	\label{energyeq}
	\frac{1}{2} \left ( \frac{4}{3} \pi \left( \frac{D_{dis}}{2} \right)^{3} \rho \right) V^{2} = \frac{4}{3} \pi \left( \frac{D}{2} \right)^{3} \rho Q_{D}^{*}
\end{equation}

\noindent where the average collision velocity $V$ is assumed to be independent of size.  This yields the relation

\begin{equation}
	\label{rdis1}
	D_{dis} = \left( \frac{2 Q^{*}_{D}}{V^{2}} \right)^{\frac{1}{3}} D.
\end{equation}

\noindent Here we consider the case where the impact strength $Q_{D}^{*}$ is given by a power law:

\begin{equation}
	\label{Qeq}
	Q^{*}_{D} = Q_{o} D^{s}, 
\end{equation}

\noindent Where $Q_o$ is a normalization constant and $s$ is the slope of Eq.~\ref{Qeq} on a log-log plot.  Inserting $Q^*_D$ from Eq.~\ref{Qeq} into Eq.~\ref{rdis1} gives

\begin{equation}
	\label{rdis2}
	D_{dis} = k_{dis} D^{\frac{3+s}{3}},
\end{equation}

\noindent where the constant $k_{dis}$ depends on fixed parameters:

\begin{equation}
	\label{kdis}
	k_{dis} = \left( \frac{2 Q_{o}}{V^2} \right)^{\frac{1}{3}}.
\end{equation}

The rate of destruction of targets of a given size range $[D,D+\diffd D]$ is proportional to the number of disruptors (bodies larger than $D_{dis}$) times the cross sectional area of the target ($\propto D^2$) times the number of targets $\diffd N$

\begin{equation}
	\label{killrate1}
	\left( \frac{\diffd (\diffd N)}{\diffd t} \right)_{dest} \propto - N_{>D_{dis}} D^{2} \diffd N.
\end{equation}

\noindent The number of disruptors (bodies greater than or equal to the minimum disruptor size) is given by integrating over the population (Eq.~\ref{popeq}) for all bodies with a diameter equal to or larger than $D_{dis}$ (given by Eq.~\ref{rdis2}).  Because $D_{dis}$ is always much smaller than the target, hence much smaller than the largest asteroid, we can integrate to $\infty$ without significantly affecting the result, so long as $p > 1$ (i.e.~there are more smaller bodies than larger bodies), which is always the case:


\begin{equation}
	\label{nkill}
	N_{>D_{dis}} = \int_{D_{dis}}^{\infty} B D'^{-p} \diffd D' = \frac{B \left( k_{dis} D^{\frac{3+s}{3}} \right)^{1-p}}{p-1}.
\end{equation}

\noindent Inserting Eq.~\ref{nkill} into Eq.~\ref{killrate1} gives

\begin{equation}
	\label{killrate2}
	\left( \frac{\diffd (\diffd N)}{\diffd t} \right)_{dest} \propto - D^{\frac{3+s}{3}(1-p)} D^{2} \diffd N \propto - D^{\frac{9 - ps - 3p + s}{3}} \diffd N,
\end{equation}

\noindent which can be written as

\begin{equation}
	\label{killrate3}
	\left( \frac{\diffd (\diffd N)}{\diffd t} \right)_{dest} = -\frac{\diffd N}{\tau},
\end{equation}

\noindent where $\tau$ is the mean collisional lifetime of bodies of diameter $D$:

\begin{equation}
	\label{lifetime}
	\tau = K D^{\frac{3p - s + ps - 9}{3}}
\end{equation}

\noindent and $K$ is a constant of proportionality.  Inserting $\tau$ and $\diffd N$ (from Eqns.~\ref{lifetime} and \ref{popeq}) into the right hand side of Eq.~\ref{killrate3} gives the collisional removal rate for bodies of diameter $D$:

\begin{equation}
	\label{killrate4}
	\left( \frac{\diffd (\diffd N)}{\diffd t} \right)_{dest} = - \frac{B D^{-p} \diffd D}{K D^{\frac{3p - s + ps - 9}{3}}} = -\frac{B}{K} D^{3 - 2p - \frac{s(p-1)}{3}} \diffd D.
\end{equation}	

\subsection{Collisional Production}

The size distribution of the fragments produced by the catastrophic disruption of a target of size $D_{o}$ can be described by a power law of similar form to Eq.~\ref{popeq} but with a different power-law index $q$ \citep{1978Icar...35....1G,1989aste.conf..778G}:

\begin{equation}
	\label{fragpop}
	\diffd N = C D^{-q} \diffd D. 
\end{equation} 

\noindent The actual size distribution of fragments may be more complex than this \citep{1993CeMDA..57....1P}, but Eq.~\ref{fragpop} is a fairly good approximation, especially for impacts between bodies that are small enough that the gravitational reaccumulation of fragments is negligible. If the diameter of the largest fragment is $b D_{o}$ (where $b$ is related to the fractional mass of the largest fragment $f_l$ by $b=f_l^{1/3}$), the normalization constant $C$ is found by requiring the cumulative number of fragments equal to or larger than $b D_{o}$ to be 1:

\begin{equation}
	\label{ceq1}
	\int_{b D_{o}}^{\infty} C D^{-q} \diffd D = 1,
\end{equation}

\noindent which yields

\begin{equation}
	\label{ceq2}
	C = (q-1)(b D_{o})^{q-1}.
\end{equation}

Eq.~\ref{ceq2} shows that the value of the coefficient $C$ depends strongly on the size of the largest fragment $bD_o$, and hence on $b$.  What value of $b$ should we adopt?  Impacts with barely enough energy to disrupt the target yield $b=(1/2)^{1/3}$.  Because the impactor population is a steep power law, most impactors are not much larger than the critical size $D_{dis}$, but the population of impactors does include larger projectiles that can yield debris with much smaller $b$ for the same target size.  Thus the effective `typical' value of $b$ for any target size is difficult to determine.  Here we simply assume that whatever the effective value of $b$ is, it is independent of $D_o$.  We will ultimately show (Section 2.3) that our results are independent of the actual value of the effective $b$, as long as $b$ is independent of $D_o$.

The relation between the power-law index $q$ of the fragment size distribution and the fractional size $b$ of the largest fragment is found by equating the volume of fragments to the volume of the parent body, where the number of fragments $\diffd N$ of diameter $D$ is given by Eq.~\ref{fragpop}.  The minimum diameter of the fragments is taken to be zero, and the maximum diameter is $b D_{o}$, which is by definition the largest fragment produced.  We assume here that $q$ is less than 4 so that the total mass of the fragments is finite even though we integrate from $D=0$.  Thus, 

\begin{align}
	\label{beq1}
	\frac{4}{3} \pi \left( \frac{D_o}{2} \right)^{3} & = \int_{0}^{b D_{o}} \frac{4}{3} \pi \left( \frac{D}{2} \right)^{3} \diffd N \nonumber \\
	& = \int_{0}^{b D_{o}} \frac{4}{3} \pi \left( \frac{D}{2} \right)^{3} (q-1)(b D_{o})^{q-1} D^{-q} \diffd D. 
\end{align}

\noindent Integration yields

\begin{equation}
	\label{beq2}
	(q-1)(b D_{o})^{q-1} \frac{(b D_{o})^{4-q}}{4-q} = D_{o}^{3}.
\end{equation}

\noindent $D_{o}$ cancels out in Eq.~\ref{beq2}.  Solving Eq.~\ref{beq2} for q gives:

\begin{equation}
	\label{qeq}
	q = \frac{b^{3}+4}{b^{3}+1},
\end{equation}

\noindent just as found by \citet{1989aste.conf..778G}. 

The production rate of fragments of a given diameter $D$ due to the breakup of bodies of diameter $D_{o}$ is given by the product of the number of fragments of diameter $D$ produced by the breakup of a body of diameter $D_{o}$ (Eq.~\ref{fragpop}, with $C$ from Eq.~\ref{ceq2}) and the breakup rate of bodies of size $D_{o}$ (Eq.~\ref{killrate4})

\begin{equation}
	\label{prodrate2}
	\left( \frac{\diffd (\diffd N)}{\diffd t} \right)_{prod} = (q-1)(b D_{o})^{q-1} D^{-q} \diffd D \frac{B}{K} D_o^{3 - 2p - \frac{s(p-1)}{3}} \diffd D_o.
\end{equation}

\noindent The total production rate is then given by integrating Eq.~\ref{prodrate2} over all $D_{o}$ which can produce fragments of size $D$.  Since the largest fragment is given by $b D_{o}$, the smallest parent body capable of producing a fragment of diameter $D$ has a diameter of $D/b$, so that this is the lower limit on the integration.  It can be shown (using the relationship which will be given in Eq.~\ref{solution1}) that the exponent of $D_o$ in the following integral is less than -1 for all $q$ less than 4.  Since $q$ must be less than 4 to prevent the total fragment mass from becoming infinite, this integral will always be finite and setting the upper integration limit to $\infty$ is essentially equivalent to setting it to the diameter of the largest body, as long as most bodies are much smaller than the largest body.  Thus, 

\begin{eqnarray}
	\label{prodrate5}
	\left( \frac{\diffd (\diffd N)}{\diffd t} \right)_{prod} & = & (q-1) \frac{B}{K} b^{q-1} D^{-q} \diffd D \int_{\frac{D}{b}}^{\infty} D_{o}^{q + 2 - 2p - \frac{s(p-1)}{3}} \diffd D_{o} \nonumber \\
	& = & \frac{B}{K} \frac{(1-q)}{q + 3 - 2p - \frac{s(p-1)}{3}} b^{2p - 4 + \frac{s(p-1)}{3}} D^{3 - 2p - \frac{s(p-1)}{3}} \diffd D.
\end{eqnarray}

\subsection{Collisional Equilibrium}

In steady-state, the rate of destruction (Eq.~\ref{killrate4}) of bodies of diameter $D$ matches the rate of production (Eq.~\ref{prodrate5}) of bodies of diameter $D$:

\begin{equation}
	\label{equil2}
	\frac{B}{K} D^{3 - 2p - \frac{s(p-1)}{3}} = \frac{B}{K} \frac{(1-q)}{q + 3 - 2p - \frac{s(p-1)}{3}} b^{2p - 4 + \frac{s(p-1)}{3}} D^{3 - 2p - \frac{s(p-1)}{3}}.
\end{equation}

\noindent $B$, $K$, and $D$ cancel, leaving

\begin{equation}
	\label{equil3}
	\frac{(1-q)}{q + 3 - 2p - \frac{s(p-1)}{3}} b^{2p - 4 + \frac{s(p-1)}{3}} = 1.
\end{equation}

\noindent Rearranging Eq.~\ref{equil3} to put $p$ on the left (except where it appears in an exponent), and using Eq.~\ref{qeq} to eliminate $q$ gives:

\begin{equation}
	\label{equil5}
	p = \frac{3}{2+\frac{s}{3}} + \frac{\frac{b^{3}+4}{b^{3}+1}}{2+\frac{s}{3}} + \frac{s}{6+s} + \left( \frac{b^{3}+4}{b^{3}+1}-1 \right) \frac{b^{2p - 4 + \frac{s(p-1)}{3}}}{2+\frac{s}{3}}.
\end{equation}

\noindent The form of this equation precludes standard algebraic solution techniques for $p$, which is the steady-state power-law index of the population.  However, by inspection we find that Eq.~\ref{equil5} is satisfied by

\begin{equation}
	\label{solution1}
	p = \frac{7 + \frac{s}{3}}{2 + \frac{s}{3}}.
\end{equation}

The population index $p$ is independent of $b$, indicating that the exact nature of the fragmentation process has little effect on the final collisionally evolved population.  For $s=0$, which corresponds to size-independent strength $Q^*_D$, this gives the classical Dohnanyi steady-state solution of $p=3.5$.    However, if the strength $Q_D^*$ varies with size ($s \neq 0$), $p$ can differ significantly from the Dohnanyi equilibrium result, with $p>3.5$ for $s<0$ and $p<3.5$ for $s>0$.  We show in Appendix A that Eq.~\ref{solution1} is a unique solution to Eq.~\ref{equil5}.

\section{Two-Slope Collisional Model}

So far we have only considered the case where a single power law describes the strength $Q_D^*$.  However, over the size range of the asteroid belt and Kuiper belt (1000 km down to sub-meter sizes), material strength is generally believed to be controlled by two different effects, depending on the portion of the size range.  The strength of asteroids less than $\sim 1$ km in diameter is controlled by the material properties such as the flaw distribution within the material \citep{1982Icar...52..409F, 1990Icar...84..226H, 1994P&SS...42.1067H, 1992PhDT.........4R, 1999Icar..142...21H, 1999Icar..142....5B}.  We refer to this portion of the population as the `strength-scaled regime.'  For asteroids larger than this, gravity dominates the effective strength of the material by gravitational self-compression and by causing the reaccumulation of collisional fragments \citep{1985Icar...63...30D, 1993CeMDA..57....1P, 1994P&SS...42.1099C, 1994P&SS...42.1067H, 1996Icar..124..141L, 1997Icar..129..562M, 1999Icar..142....5B}.  We refer to this portion of the population as the `gravity-scaled regime.'  In these two different regimes, two different power laws (Eq.~\ref{Qeq}) can approximate the $Q_D^*$ vs.~size relationship, each with its own slope $s$.  In the strength-scaled regime, $s$ is negative and in the gravity-scaled regime, $s$ is positive.  The various power-law estimates of $Q^*_D$ by the aforementioned authors are shown in Fig.~\ref{qrange}.

In this section, we address the expected size distribution for a population with strength $Q_D^*$ following one power law with slope $s_s$ (`strength-scaled') for bodies smaller than diameter $D_t$, connected to another power law with slope $s_g$ (`gravity-scaled') for large bodies as shown in Fig.~\ref{qfig}.

\subsection{Gravity Scaled Portion of the Population}

The gravity scaled portion of the population ($D>D_t$) is fed entirely by collisional fragments from larger bodies (in the gravity-scaled regime), and except for target bodies close to the transition diameter $D_t$, they are destroyed by bodies within the gravity regime as well.  Because the gravity-scaled regime is approximately `self-contained' in this manner, its population index $p_g$ is simply given by the single-slope solution (Eq.~\ref{solution1}), with subscript $g$ added to indicate the gravity-scaled regime

\begin{equation}
	\label{p_g_sol}
	p_g = \frac{7 + \frac{s_g}{3}}{2 + \frac{s_g}{3}}.	
\end{equation}    

\noindent Since $s_g$ is positive in the gravity-scaled regime, Eq.~\ref{p_g_sol} yields a population index $p_g$ less than 3.5 in the gravity-scaled regime

However, some targets in the gravity-scaled regime (those only slightly larger than $D_t$) will be destroyed by bodies in the strength-scaled regime, where the population may not follow the same power law as in the gravity-scaled regime.  Those events will lead to perturbations to the gravity-scaled population.  In Section 3.3, we show that these perturbations are wavelike oscillations about a power law with an index $p_g$ given by Eq.~\ref{p_g_sol}, and these perturbations affect neither the destruction rate of bodies in the gravity-scaled regime nor the production rate of bodies smaller than $D_t$ by bodies in the gravity-scaled regime.  Therefore, even though the population in the gravity-scaled regime may be perturbed from a strict power law of index $p_g$, it still follows the general trend of a power law of index $p_g$ and it still behaves (in terms of collisional production and destruction) as if it were a power-law size distribution with index $p_g$.

\subsection{Strength Scaled Portion of the Population}

The strength-scaled portion of the population ($D<D_t$) is broken up almost exclusively by bodies within the strength-scaled regime, since the minimum sized impactor for disruption is generally much smaller than the target body in the asteroid belt and there are many more small bodies than large ones.  However, the production of new bodies in the strength-scaled regime is due to the fragmentation of larger bodies in both the strength- and gravity-scaled regimes.  We assume that strength-scaled portion of the population follows a power law with index $p_s$, which is likely to be different from the population index $p_g$ in the gravity-scaled regime.  Thus, we must explicitly account for the contribution of collisional fragments from both regimes.  Equation \ref{prodrate5} for the production rate of fragments in the single-slope case can be extended to treat a population described by 2 power laws:

\begin{eqnarray}
	\label{twoslopeprod1}
	\left( \frac{\diffd (\diffd N)}{\diffd t} \right)_{prod} = (q-1) \frac{B_s}{K_s} b^{q-1} D^{-q} \diffd D \int_{\frac{D}{b}}^{D_t} D_{o}^{q + 2 - 2p_s - \frac{s_s(p_s-1)}{3}} \diffd D_{o} + \nonumber \\ (q-1) \frac{B_g}{K_g} b^{q-1} D^{-q} \diffd D \int_{D_t}^{\infty} D_{o}^{q + 2 - 2p_g - \frac{s_g(p_g-1)}{3}} \diffd D_{o}.
\end{eqnarray}

\noindent Here we ignore the deviation from a power law among bodies with $D>D_t$, which as noted in Section 3.1 will have a negligible effect on the production rate of bodies with $D<D_t$.  Eq.~\ref{twoslopeprod1} can be integrated and rearranged to give

\begin{eqnarray}
	\label{near_final1}
	\left( \frac{\diffd (\diffd N)}{\diffd t} \right)_{prod} = \frac{B_s}{K_s} \frac{(1-q)}{q + 3 - 2p_s - \frac{s_s(p_s-1)}{3}} b^{2p_s - 4 + \frac{s_s(p_s-1)}{3}} D^{3 - 2p_s - \frac{s_s(p_s-1)}{3}} \diffd D + \nonumber \\ (1-q) b^{q-1} \left( \frac{D_t}{D} \right)^q \left[ \frac{B_g}{K_g} \frac{D_t^{3 - 2 p_g - \frac{s_g(p_g-1)}{3}}}{q + 3 - 2p_g - \frac{s_g(p_g-1)}{3}} - \frac{B_s}{K_s} \frac{D_t^{3 - 2 p_s - \frac{s_s(p_s-1)}{3}}}{q + 3 - 2p_s - \frac{s_s(p_s-1)}{3}} \right] \diffd D.
\end{eqnarray}

A relation between $B_s$ and $B_g$ can be derived from the fact that the numbers of bodies must match at $D_t$.  Following Eq.~\ref{popeq},  

\begin{equation}
	\label{bmatch}
	B_g D_t^{-p_g} = B_s D_t^{-p_s}.
\end{equation}

\noindent Similarly, the relation between $K_s$ and $K_g$ can be found by equating the lifetimes (Eq.~\ref{lifetime}) at $D_t$  

\begin{equation}
	\label{kmatch}
	K_g D_t^{\frac{3 p_g - s_g + p_g s_g - 9}{3}} = K_s D_t^{\frac{3 p_s - s_s + p_s s_s - 9}{3}}.
\end{equation}

\noindent Combining Eqns.~\ref{bmatch} and \ref{kmatch} gives

\begin{equation}
	\label{bkrelate}
	\frac{B_g}{K_g} = \frac{B_s}{K_s} D_t^{2(p_g - p_s) - \frac{s_s p_s - s_g p_g}{3} - \frac{s_g - s_s}{3}}.
\end{equation}

\noindent Substituting Eq.~\ref{bkrelate} into Eq.~\ref{near_final1} gives

\begin{eqnarray}
	\label{newprodrate}
	\left( \frac{\diffd (\diffd N)}{\diffd t} \right)_{prod} = \frac{B_s}{K_s} \frac{(1-q)}{q + 3 - 2p_s - \frac{s_s(p_s-1)}{3}} b^{2p_s - 4 + \frac{s_s(p_s-1)}{3}} D^{3 - 2p_s - \frac{s_s(p_s-1)}{3}} \diffd D + \nonumber \\ (1-q) b^{q-1} \left( \frac{D_t}{D} \right)^q \frac{B_s}{K_s} D_t^{3 - 2 p_s - \frac{s_s(p_s-1)}{3}} \left[ \frac{1}{q + 3 - 2p_g - \frac{s_g(p_g-1)}{3}} - \frac{1}{q + 3 - 2p_s - \frac{s_s(p_s-1)}{3}} \right] \diffd D.	
\end{eqnarray}

\noindent In a steady state, the destruction rate must match the production rate.  Equating the destruction rate of bodies in the strength-scaled regime (given by Eq.~\ref{killrate4}, with proper subscripts to indicate that all projectiles and targets are in the strength-scaled regime) with the production rate from Eq.~\ref{newprodrate} (by bodies in both the gravity-scaled and strength-scaled regimes) yields

\begin{align}
	\label{strengthslope}
	\frac{B_s}{K_s} D^{3 - 2p_s - \frac{s_s(p_s-1)}{3}} = 
	\frac{B_s}{K_s} \frac{(1-q)}{q + 3 - 2p_s - \frac{s_s(p_s-1)}{3}}  b^{2p_s - 4 + \frac{s_s(p_s-1)}{3}} D^{3 - 2p_s - \frac{s_s(p_s-1)}{3}} +  \nonumber \\ (1-q) b^{q-1} \left( \frac{D_t}{D} \right)^q \frac{B_s}{K_s} D_t^{3 - 2 p_s - \frac{s_s(p_s-1)}{3}} \left[ \frac{1}{q + 3 - 2p_g - \frac{s_g(p_g-1)}{3}} - \frac{1}{q + 3 - 2p_s - \frac{s_s(p_s-1)}{3}} \right].   
\end{align}  

\noindent Note that the left hand side and the first term on the right hand side of Eq.~\ref{strengthslope} have the same form as the single-slope expression (Eq.~\ref{equil2}).  The last term depends on both $p_s$ and $p_g$, and  accounts for the fact that the population index $p_g$ in the gravity-scaled regime (given by Eq.~\ref{p_g_sol}) likely differs from the population index $p_s$ in the strength-scaled regime, and this may affect the production rate.  Eq.~\ref{strengthslope} relates $p_s$, the slope of the population smaller than $D_t$, to known quantities.  The form of Eq.~\ref{strengthslope} precludes a simple algebraic solution.  However, note that when we insert $p_g$ from Eq.~\ref{p_g_sol}, the last term goes to zero if

\begin{equation}
	\label{p_s_sol}
	p_s = \frac{7 + \frac{s_s}{3}}{2 + \frac{s_s}{3}}.	
\end{equation}

\noindent Moreover, this solution makes the left hand side of Eq.~\ref{strengthslope} match the first term on the right hand side, solving Eq.~\ref{strengthslope} entirely.  Since $s_s$ is negative in the strength-scaled regime, Eq.~\ref{p_s_sol} yields a population index $p_s$ greater than 3.5 in the strength-scaled regime.

This result shows that the population index in the strength-scaled regime (Eq.~\ref{p_s_sol}) is completely independent of the population index and the slope $s_g$ of $Q^*_D$ in the gravity-scaled regime, and indeed has the same form as the solution for a case where strength $Q_D^*$ follows a single power law (Eq.~\ref{solution1}).  Evidently, the production rate of bodies in a colliding population is unaffected by a change in power-law index of that population at larger sizes.  This result can be understood as follows: Even though the population index in the gravity-scaled regime is smaller (i.e.~the population has a shallower slope) than in the strength-scaled regime and would therefore contain a relatively larger number of bodies than would be predicted if strength-scaling continued to large sizes, the increased strength of bodies in the gravity-scaled regime exactly offsets these increased numbers, such that the total breakup rate of bodies of a given size (and hence the fragment production rate by bodies of a given size) is not affected.  For that reason, the last term in Eq.~\ref{strengthslope} goes to zero and the solution for $p_s$ (Eq.~\ref{p_s_sol}) is independent of $p_g$ and $s_g$.

\subsection{Waves in the Size Distribution}

The population indices of the two portions of the size distribution (larger and smaller than $D_t$) are independent of one another, as shown in Sections 3.1 and 3.2 respectively.  In each regime, the index $p$ depends only on the slope $s$ of the $Q_D^*$ law describing the strength for that size range. However, as noted in Section 3.1, effects of the transition in strength and population index near $D_t$ introduces some deviation from a strict power law for bodies larger than $D_t$.  In this section, we quantify this deviation and show that it does not affect the general population index $p_g$ in this regime, nor does it have significant effect on the population index for smaller bodies.

In the derivation of the population index $p_g$ in the gravity-scaled regime (Section 3.1), we assumed that all asteroids were disrupted by projectiles whose numbers were described by the same power law.  However, for those targets just larger than $D_t$ (i.e.~near the small end of the gravity-scaled regime), projectiles are mostly smaller than $D_t$, and hence are governed by the strength-scaled size distribution.  Consider the two steady-state power laws describing the population in the strength- and gravity-scaled regimes, joined at the transition diameter $D_{t}$ (Fig.~\ref{wavefig}a).  Let $D_{t_{dis}}$ be the diameter of the body which can disrupt a body of diameter $D_{t}$.  Due to the transition from the gravity-scaled regime to the strength-scaled regime below $D_{t}$, bodies of diameter $D_{t_{dis}}$ are more numerous than would be expected by assuming that all bodies are gravity scaled, leading to a configuration that is not in a steady state.

A steady-state configuration can be achieved by `sliding' the population in the strength-scaled regime down in number, as shown in Fig.~\ref{wavefig}b. To determine the magnitude of the shift, consider the production and destruction rates of bodies of diameter $D_{t}$.  Since the destruction rate is proportional to the number of projectiles (Eq.~\ref{killrate1}), an excess $\Delta \mathrm{Log} N(D_{t_{dis}})$ of bodies of diameter $D_{t_{dis}}$ which are capable of destroying bodies of diameter $D_t$ causes a proportional increase in the destruction rate of bodies of diameter $D_t$.  However, since the destruction rate is also proportional to the number of targets (Eq.~\ref{killrate1}), a decrease $\Delta \mathrm{Log} N(D_t)$ of bodies of diameter $D_t$ results in a proportional decrease in the destruction rate of bodies of diameter $D_t$.  We assume here that the production rate of bodies of diameter $D_t$ stays constant, despite these changes (this was noted in Section 3.1, and will be shown to be true later in this section).  In a steady-state, the destruction rate of bodies of diameter $D_t$ is equal to the production rate.  With the excess $\Delta \mathrm{Log} N(D_{t_{dis}})$ of projectiles and the depletion $\Delta \mathrm{Log} N(D_{t})$ of targets, the destruction rate of bodies of diameter $D_t$ is related to the production rate by

\begin{equation}
	\mathrm{Log} \left( \frac{\diffd (\diffd N (D_t))}{\diffd t} \right)_{dest} = \mathrm{Log} \left( \frac{\diffd (\diffd N (D_t))}{\diffd t} \right)_{prod} + \Delta \mathrm{Log} N(D_{t_{dis}}) + \Delta \mathrm{Log} N(D_t).
\end{equation}

\noindent From the previous equation, the production and destruction rates are equal (and the system is in a true steady state) when $\Delta \mathrm{Log} N(D_{t_{dis}})$ and $\Delta \mathrm{Log} N(D_t)$ are equal and opposite, as shown in Fig.~\ref{wavefig}b.  To calculate the magnitude of the shift $\Delta \mathrm{Log} N(D_t)$, we first find the logarithmic difference between $D_{t}$ and $D_{t_{dis}}$ (using Eq.~\ref{rdis1}):

\begin{equation}
	\label{D_diff}
	\Delta \mathrm{Log} D = \mathrm{Log} D_{t} - \mathrm{Log} D_{t_{dis}} =  -\frac{1}{3} \mathrm{Log} \left(\frac{2 Q_{t}}{V^{2}}\right),
\end{equation}

\noindent where $Q_t = Q^*_D(D_t)$.  Using Eq.~\ref{D_diff} and the difference between the two population indices $p_s$ and $p_g$, we calculate 

\begin{equation}
	\label{wave_amplitude}
	\Delta \mathrm{Log} N(D_t) = \frac{1}{6} \mathrm{Log} \left(\frac{2Q_{t}}{V^{2}}\right) (p_{s} - p_{g}).
\end{equation}

In reality, the shift in number of bodies at $D_t$ given by Eq.~\ref{wave_amplitude} does not result in a simple discontinuity as shown in Fig.~\ref{wavefig}b, but instead causes perturbations to the size distribution in the gravity-scaled regime ($D>D_t$).  The underabundance $\Delta \mathrm{Log} N(D_t)$ of bodies of diameter $D_t$ (a `valley') leads to an overabundance of bodies which impactors of diameter $D_{t}$ are capable of destroying (a `peak'), which in turn leads to another `valley' and so on.  This results in a wave of amplitude $\left| \Delta \mathrm{Log} N(D_t) \right|$ which propagates through the large body size distribution as shown in Fig.~\ref{wavefig}c. 

The average power-law index $p_g$ of the population in the gravity-scaled regime will not be significantly changed by the initiation of this wave.  A `peak' in the wave will have $\left| \Delta \mathrm{Log} N(D_t) \right|$ more bodies relative to a straight power law, but the following valley will have $\left| \Delta \mathrm{Log} N(D_t) \right|$ fewer bodies than a straight power law and so on---the wave oscillates about a power law of slope $p_g$ (given by Eq.~\ref{p_g_sol}).  Likewise, the production rate of smaller bodies ($D<D_t$) by bodies in the gravity-scaled regime will not be significantly affected by the wave.  While a `peak' in the wave may have $\left| \Delta \mathrm{Log} N(D_t) \right|$ more target bodies than expected from the power-law case, the bodies capable of disrupting those target bodies lie in a `valley' and are hence lower in number by a factor of $\left| \Delta \mathrm{Log} N(D_t) \right|$.  The increase in number of targets is offset by a decrease in the number of impactors, such that the breakup rate and hence the fragment production rate is unchanged.  The opposite is true as well.  Bodies which lie in a `valley' are catastrophically disrupted by bodies at a `peak'.  The end result is that the waviness induced in the gravity-scaled portion of the size distribution does not change the production rate of smaller bodies.

The position of the peaks and valleys of the wave can be found from Eqns.~\ref{rdis2} and \ref{kdis}.  Equation \ref{kdis} is written in terms of the normalization constant $Q_o$, which in this case is the value $Q_D^*$ would have at 1 m or 1 km (depending on the units used for diameter) if it followed a single power law of slope $s_g$.  Using Eq.~\ref{Qeq}, $Q_o$ can be written in terms of the more intuitive values $Q_t$ and $D_t$ as $Q_o = Q_t D_t^{-s_g}$.  Given the diameter $D_{v}$ where there is a valley (such as at $D_{t}$), the peak which follows will be at the diameter $D_{p}$: 

\begin{equation}
	\label{dp}
	D_{p} = \left(\frac{2 Q_{t}}{V^{2}}\right)^{-\frac{1}{3+s_g}} D_t^{\frac{s_g}{3+s_g}} D_{v}^{\frac{3}{3+s_g}}.
\end{equation}  

\noindent Conversely, when the diameter $D_{p}$ where a peak occurs is known, the diameter $D_{v}$ of the valley which follows is given by 

\begin{equation}
	\label{dv}
	D_{v} = \left(\frac{2 Q_{t}}{V^{2}}\right)^{-\frac{1}{3+s_g}} D_t^{\frac{s_g}{3+s_g}} D_{p}^{\frac{3}{3+s_g}}.
\end{equation}  

Thus, with Eqns.~\ref{p_g_sol} and \ref{p_s_sol} we can find the power law indices which describe the strength- and gravity-scaled portions of the population, and with Equations \ref{wave_amplitude}, \ref{dp}, and \ref{dv} we can quantify the `wavy' structure which is superimposed on the power law describing the gravity-scaled portion of the population. 

\section{Comparison to Numerical Results}

\noindent In this section we compare our analytical results with numerical collisional evolution simulations.  Our numerical model tracks a population binned in logarithmic intervals $\mathrm{d} \mathrm{Log} D$ in diameter.  The number of bodies in each bin is $\mathrm{d} N$.  At each timestep, the minimum disruptor diameter $D_{dis}$ for bodies in each bin is calculated by Eq.~\ref{rdis1}, assuming a $Q^*_D$ law and a mean collision velocity (we use $V=5300$ m/s, from \citet{1993GeoRL..20..879B} and \citet{1994Icar..107..255B}), but the results are not substantially different for values $20\%$ larger or smaller than this).  The lifetime of bodies in each bin is calculated by

\begin{equation}
	\tau = \frac{4}{N(>D_{dis}) D^{2} P_{i}},
\end{equation}

\noindent where $P_i$ is the intrinsic collisional probability, which is approximately $3 \times 10^{-18} \ \mathrm{km^{-2} \ yr}$ for the asteroid belt \citep{1992Icar...97..111F, 1993GeoRL..20..879B, 1994Icar..107..255B, 1997Icar..130..140D}, and $N(>D_{dis})$ is calculated by summing over all bins equal in size to or larger than $D_{dis}$.  The removal rate of bodies from each bin is then calculated from Eq.~\ref{killrate3}, and this is used to calculate the number of bodies removed from each bin during each timestep.  For each body removed by catastrophic disruption, fragments are produced according to the distribution given by Eqns.~\ref{fragpop} and \ref{ceq2}, and this is used to determine the number of new fragments added to each bin during each timestep.  Cratering debris are neglected in this model.

We first performed a series of numerical simulations to determine the variation of population index with the slope of the scaling law $Q^*_D$.  We use an initial population with index 3.5 (the predicted Dohnanyi equilibrium value), evolve it it time for 4.5 Gyr with strength law slope $s$ ranging from -1 to 2, and measure the final power-law index $p$ of the population. Fig.~\ref{numcomp_fig} shows the excellent agreement between our numerical results and the analytical relation given in Eq.~\ref{solution1}.  We performed similar simulations with initial populations having different power-law indices, and found that the results are essentially the same as those shown in Fig.~\ref{numcomp_fig} (i.~e.~the final population index does not depend on the starting index).  \citet{durda:thesis} presented a figure similar to Fig.~\ref{numcomp_fig}, derived entirely from numerical results, for $s$ from -0.3 to 0.3, and a slightly revised figure appeared in \citet{1997Icar..130..140D} in which a ``second-order effect'' (i.e.~a somewhat nonlinear relationship between $p$ and $s$) was noted.  Our more extensive numerical results match those of \citet{durda:thesis} and \citet{1997Icar..130..140D}, and our analytical results confirm and explain their numerical results, including the second-order effects noted in \citet{1997Icar..130..140D}.

Next, we performed a series of simulations with jointed power laws describing the strength law $Q^*_D$ in order to test our analytical predictions.   Fig.~\ref{sim1} shows the results of such a simulation, comparing the initial population (with a power-law index of $3.5$) and the final population 4.5 Gyr later.  The curve labeled `Strength+Gravity Scaling' is evolved with 2 power laws fit to the \citet{1999Icar..142....5B} $Q^*_D$ scaling law ($s_s=-0.36$, $s_g=1.36$, $D_t=0.7$ km, and $Q_t=200 \ \mathrm{J/kg}$) (see Fig.~\ref{q_num}), while the curve labeled `All Gravity Scaled' has been evolved assuming that even for bodies smaller than $D_t$, $Q^*_D$ follows a power law with $s_s=s_g=1.36$.  Note the waves which appear in the large body population when there is a transition in the slope of the $Q_D^*$ scaling law.  The population is binned and plotted in logarithmic intervals $\mathrm{d} \mathrm{Log} D$, so the slopes on the plot are 1 larger than the exponent $-p$ in Eq.~\ref{popeq}.  Thus, for example, the Dohnanyi equilibrium power-law index $p=3.5$ would appear as a slope of $-2.5$ if plotted in Fig.~\ref{sim1}.  When referring to our plot, we give the value of $p$ to facilitate comparison to our analytical predictions.

The population evolved with pure gravity scaling has $p_g=3.04$, which is exactly the value predicted by Eq.~\ref{p_g_sol} for $s_g=-1.36$.  However, there is some deviation from this slope for bodies larger that $\sim$10 km due to the fact that above $\sim$10 km, the mean collisional lifetimes of bodies begin to approach or exceed $4.5$ Gyr and hence the population does not have enough time to reach a steady state.  Running the simulation for longer times (20 Gyr and 100 Gyr) decreases this deviation but does not completely eliminate it, indicating that even after 100 Gyr, the largest bodies are not fully collisionally relaxed.

For the population evolved with both strength and gravity scaling, the index of the strength-scaled portion of the population ($D<D_t$) is $p_s=3.66$, which is exactly the value predicted by Eq.~\ref{p_s_sol} for $s_s=0.36$.  The gravity-scaled portion of the population ($D>D_t$) is wavy, and oscillates about the population evolved with pure gravity scaling ($p_g=3.04$).  This example, along with other simulations we have performed, confirms our predictions that: (1) The population index in the strength-scaled regime is independent of the population index and $Q_D^*$ law in the gravity-scaled regime, and (2) While the gravity-scaled portion of the population is wavy, it follows the general trend of a power law with index $p_g$ that is independent of the population index and $Q_D^*$ law in the strength-scaled regime.

The amplitude of the wave $\left| \Delta \mathrm{Log} N(D_t) \right|$ in the simulation is found to be $0.48$, which is very close to the value of $0.50$ from Eq.~\ref{wave_amplitude}.  The wave amplitude decreases slightly for the following peak and valley, and there is effectively no peak formed around 100 km because the collisional lifetime of 100 km bodies is so large and there are few larger bodies to resupply new 100 km bodies by collisions.  Arrows on Fig.~\ref{sim1} show the positions of the peaks and valleys predicted from Eqns.~\ref{dp} and \ref{dv}.  The predicted positions overestimate the actual positions by about 30 percent.  

When we run longer numerical simulations (20 Gyr and 100 Gyr), we find that there is still some variation in the amplitudes of the peaks and valleys (i.e.~they are not all the same as $\left| \Delta \mathrm{Log} N(D_t) \right|$).  In addition, there is still some discrepancy between the predicted positions of the peaks and valleys and the actual positions.  This is due in part to the fact that even after 100 Gyr, the largest bodies may not be fully collisionally relaxed.  In addition, numerical modeling can more accurately simulate the collisional evolution process, uncovering second-order effects which our analytical model does not account for.

\section{Summary and Implications}

\noindent We have analytically derived the steady-state power-law index of a collisional cascade in which the material strength varies as a function of size. Earlier work had only treated collisions which are entirely self-similar, in which every body has the same strength per unit mass.  Our results are applicable to actual collisional populations, such as the asteroid belt, where material strength has been shown to be strongly size-dependent. 

For the case where a single power law describes the dependence of strength on size, we show that there is a simple analytical relation between the power-law slope $s$ of the $Q_D^*$ law describing the strength and the steady-state power-law index $p$ of the population (Eq.~\ref{solution1}).  For the self-similar case ($s=0$), our result yields the classical Dohnanyi result of $p=3.5$, but for other values of $s$, the steady-state population may have $p$ quite different from $3.5$.

For the case where large (`gravity-scaled') and small (`strength-scaled') bodies are controlled by different power-law expressions for $Q_D^*$, we find that the steady-state population index $p_s$ in the strength-scaled regime is independent of the steady-state population index $p_g$ and slope $s_g$ of $Q_D^*$ in the gravity-scaled regime and vice versa.  The steady-state population index in both regimes can be described by the same relation as in the single-slope case (Eq.~\ref{solution1}), with $p_s$ depending only on $s_s$ and $p_g$ depending only on $s_g$.  Thus, for a plausible $Q_D^*$ law with $s_s=-0.36$ for bodies smaller than $\sim$ 700 m and $s_g=1.36$ for larger bodies \citep{1999Icar..142....5B}, the population indices are $p_s=3.66$ and $p_g=3.04$.  The transition between the different population indices in the strength- and gravity-scaled regimes leads to wavelike perturbations about a power law in the gravity-scaled regime.  We have derived simple analytical expressions for the amplitude of these waves (Eq.~\ref{wave_amplitude}) and the spacing of the peaks and valleys of the wave (Eq.~\ref{dp} and \ref{dv}).  Our analytical results have been tested and validated by comparison with a numerical simulation.

Our analytical solution provides a tool for interpreting the size distribution of the asteroid belt in order to infer its strength properties.  It should be noted, however, that effects not treated in our analytical solution could potentially alter the size distribution of asteroids and lead to discrepancies between our predictions and the actual strength properties of asteroids.  For example, a small size cutoff in the size distribution due to Poynting-Robertson drag and solar radiation pressure can potentially introduce a wave in the size distribution \citep{durda:thesis,1994P&SS...42.1079C,1997Icar..130..140D}.  Such a wave would begin in the strength-scaled regime and could interfere constructively or destructively with the waves generated by the transition between strength- and gravity-scaled regimes.  The actual degree to which small particles are removed from the asteroid belt by the Poynting-Robertson effect and solar radiation pressure is not well known, and detailed analysis of such effects can only be done numerically---hence these effects are not treated here. In addition, the actual asteroid population may differ from our analytical predictions if the size distribution of collisional fragments differs significantly from the single power law assumed in Eq.~\ref{fragpop} or depends significantly on the size of the target.  These effects are difficult to model analytically, and hence are not treated here.   

A number of recent estimates of the main belt population with $D>1$ km have been published \citep{1998Icar..131..245J, 2001AJ....122.2749I}, and cratering records on asteroids such as Gaspra, Ida, Mathilde, and Eros \citep{1994Icar..107...84G, 1996Icar..120..106G, 1999Icar..140...28C, 2002Icar..155..104C} can be used to estimate the population of asteroids down to a few meters.  Thus, we have estimates of the asteroid size distribution in both the strength- and gravity-scaled regimes.  

Greenberg \textit{et al.}~(1994, 1996) found that the crater population on both Gaspra and Ida was fit best by an impacting population which had a power-law index of $p=4$ below 100 meters in diameter.  These bodies are small enough to be in the strength-scaled regime (Fig.~\ref{qrange}).  Using Eq.~\ref{p_s_sol}, we find that an index of $p=4$ for the population implies a slope of $s=-1$ for the $Q^*_D$ scaling law.  This is significantly steeper than any predictions for the strength-scaled regime shown in Fig.~\ref{qrange}, where the steepest predicted slope of the scaling law is $s=-0.61$, a value for weak mortar \citep{1992PhDT.........4R}.

For larger asteroids (3 $<$ D $<$ 30 km), \citet{1998Icar..131..245J} find that the population is very wavy and the population index varies significantly with absolute magnitude.  These bodies are large enough to be in the gravity-scaled regime (Fig.~\ref{qrange}).  Using their Tables IV and VI, we find that the average value of the population index $p_g$ is between 2.8 and 2.9, which, using Eq.~\ref{p_g_sol}, corresponds to a slope $s_g$ between 1.9 and 2.3 for the $Q_D^*$ scaling law.  The values of $s_g$ we find are consistent with the estimate of \citet{1985Icar...63...30D} but somewhat steeper than the estimates of \citet{1994P&SS...42.1067H}, \citet{1996Icar..124..141L}, \citet{1997Icar..129..562M} and \citet{1999Icar..142....5B} (see Fig.~\ref{qrange}).  Since the population in this size range is so wavy, it is possible that the average $p_g$ we use would be different if we included data for larger or smaller bodies than those treated by \citet{1998Icar..131..245J}, hence our estimate of $s_g$ could be skewed.  Likewise, if the larger bodies (those around 30 km) observed by \citet{1998Icar..131..245J} are not in collisional equilibrium, this would also affect our estimate of $s_g$.

For the NEA population, estimates are available down to around 10 m \citep{2000NatureRabinowitz}.  The NEA size distribution, however, is significantly influenced by size-dependent dynamical processes during the delivery of NEAs from the main belt.  Applying our results to the NEA population would require a full treatment of these dynamical phenomena, and we do not address that problem in this paper.

The Kuiper belt size distribution has not been well determined below about 100 km.  Moreover, the population larger than 100 km is most likely primordial and not collisionally relaxed, i.~e.~not in a steady-state \citep{1997Icar..125...50D}.  Therefore our results do not apply to that portion of the population.  With ongoing and future surveys, knowledge of the KBO size distribution will be extended to smaller sizes.  Our results will facilitate interpretation of these observations just as they do for the main belt of asteroids.

\section{Acknowledgments}

We thank Daniel D. Durda and Jean-Marc Petit for their helpful reviews and comments.  This work is supported by grants from NASA's Planetary Geology and Geophysics Program and Graduate Student Research Program.

\appendix

\section{Proof of Uniqueness of Analytical Solution}

Here we show that the relation between $p$ and $s$ derived in Sec.~2.3 is a unique solution to the equation for collisional equilibrium.  The equation for collisional equilibrium (Eq.~\ref{equil5}), repeated here, is

\begin{equation}
	\label{appendix0}
	p = \frac{3}{2+\frac{s}{3}} + \frac{\frac{b^{3}+4}{b^{3}+1}}{2+\frac{s}{3}} + \frac{s}{6+s} + \left( \frac{b^{3}+4}{b^{3}+1}-1 \right) \frac{b^{2p - 4 + \frac{s(p-1)}{3}}}{2+\frac{s}{3}}.
\end{equation}

\noindent Eq.~\ref{appendix0} has a solution (given previously in Eq.~\ref{solution1})

\begin{equation}
	\label{appendix1}
	p = \frac{7 + \frac{s}{3}}{2 + \frac{s}{3}}.
\end{equation}

\noindent Eq.~\ref{appendix0} can be rewritten by grouping the terms which do not depend on $p$ into constants $C_1$ through $C_4$

\begin{equation}
	\label{appendix2}
	p = C_1 + C_2 b^{C_3 p - C_4},
\end{equation}

\noindent where

\begin{eqnarray}
	C_1 & = & \frac{3}{2+\frac{s}{3}} + \frac{\frac{b^{3}+4}{b^{3}+1}}{2+\frac{s}{3}} + \frac{s}{6+s} \nonumber \\ 
	C_2 & = & \frac{\frac{b^{3}+4}{b^{3}+1}-1}{2+\frac{s}{3}} \nonumber \\ 
	C_3 & = & 2 + \frac{s}{3} \nonumber \\ 
	C_4 & = & -4 - \frac{s}{3}.	
\end{eqnarray}

\noindent The left hand side of Eq.~\ref{appendix2} is a monotonically increasing function of $p$.  Since $b$ (the fractional diameter of the largest fragment) is always between 0 and 1, and the constants $C_2$ and $C_3$ are positive for $s>-6$ (where $s$ is unlikely to be much less than $-1$), the right hand side is a monotonically decreasing function of $p$ (regardless of the sign of $C_1$ and $C_4$).  Since a monotonically increasing function and a monotonically decreasing function can only intersect at a single point, Eq.~\ref{appendix1} is the only solution to Eq.~\ref{appendix0}.

\clearpage


\centerline{\textbf{FIGURE CAPTIONS}}

\figcaption{Estimates of $Q^*_D$, the critical specific energy required for catastrophic disruption.  Lines with negative slope show weakening with increasing size, as expected for bodies smaller than $\sim$1 km (`strength-scaled').  Lines with positive slope show strengthening with increasing size, as expected for bodies larger than $\sim$1 km, for which self-gravity is important (`gravity-scaled').  \label{qrange}}

\figcaption{Hypothetical $Q_D^*$ law for a population with different strength properties for large and small bodies.  $Q_D^*$ consists of two different power laws with slopes $s_s$ and $s_g$ joined at the transition diameter $D_t$.  In the strength-scaled regime, material properties control the effective strength, while in the gravity-scaled regime, gravity dominates the effective strength through self-compression and gravitational reaccumulation of collisional fragments.  \label{qfig}}

\figcaption{Sequence showing how waves form in the population as a result of a change in strength properties at $D_t$.  (a) For a $Q_D^*$ law such as that shown in Fig.~\ref{qfig}, the resulting steady-state population is steeper for smaller, strength-scaled bodies (population index $p_s$) than for larger, gravity-scaled bodies (population index $p_g$).  Thus, impactors capable of destroying bodies of diameter $D_t$ are overabundant relative to what would be expected by extrapolating the gravity regime slope.  This configuration is not in collisional equilibrium. (b) To counteract this, the number of bodies of diameter $D_t$ and smaller decreases by a factor $\Delta \mathrm{Log} N(D_t)$ so that there are fewer `targets' of diameter $D_t$ and fewer impactors of diameter $D_{t_{dis}}$.  (c) The decrease in bodies of diameter $D_t$ leads to an overabundance of bodies which can be destroyed by impactors of diameter $D_t$, which in turn leads to a depletion of larger bodies and so on.  Thus, a wave is formed in the large-body population.  \label{wavefig}}

\figcaption{Comparison between our analytical relationship between the slope $s$ of the strength law $Q^*_D$ and the steady-state power-law index $p$ of population (Eq.~\ref{solution1}), and series of numerical simulations.  The analytical relationship is shown as a solid line, and the results from numerical simulations are shown as open circles.  \label{numcomp_fig}}

\figcaption{Results of a 4.5 Gyr numerical collisional evolution simulation.  The solid line is the initial population, the long-dashed line is the final population assuming all bodies have a gravity-scaled $Q_D^*$, and the short-dashed line is the final population where small bodies have a strength-scaled $Q^*_D$ and large bodies have a gravity-scaled $Q^*_D$.  For the latter case, the transition between the two different strength regimes leads to `waves' which propagate through the large-body population.  Arrows show the positions of the `peaks' and `valleys' as predicted by Eqns.~\ref{dp} and \ref{dv}.  \label{sim1}}

\figcaption{$Q^*_D$ law used in the numerical simulation presented in Fig.~\ref{sim1}, consisting of 2 power laws fit to the \citet{1999Icar..142....5B} law.  The actual \citet{1999Icar..142....5B} $Q^*_D$ is plotted for comparison.  The sharply joined, 2 power-law fit is used in the simulations to make the positions of the peaks and valleys of the wave more clear. \label{q_num}}

\clearpage


\begin{figure}
\plotone{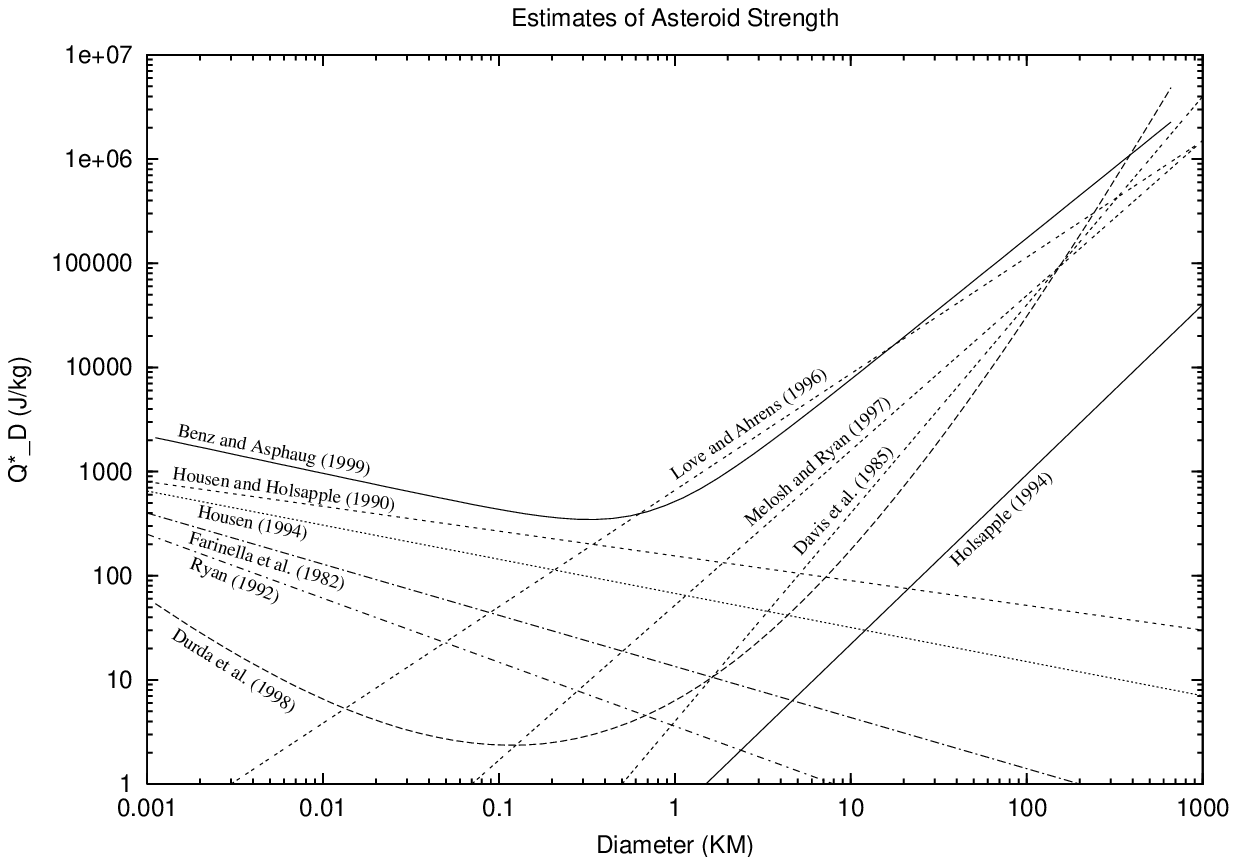}
\centerline{ }
\centerline{ }
\centerline{\Large{Figure 1}}
\end{figure}

\clearpage

\begin{figure}
\plotone{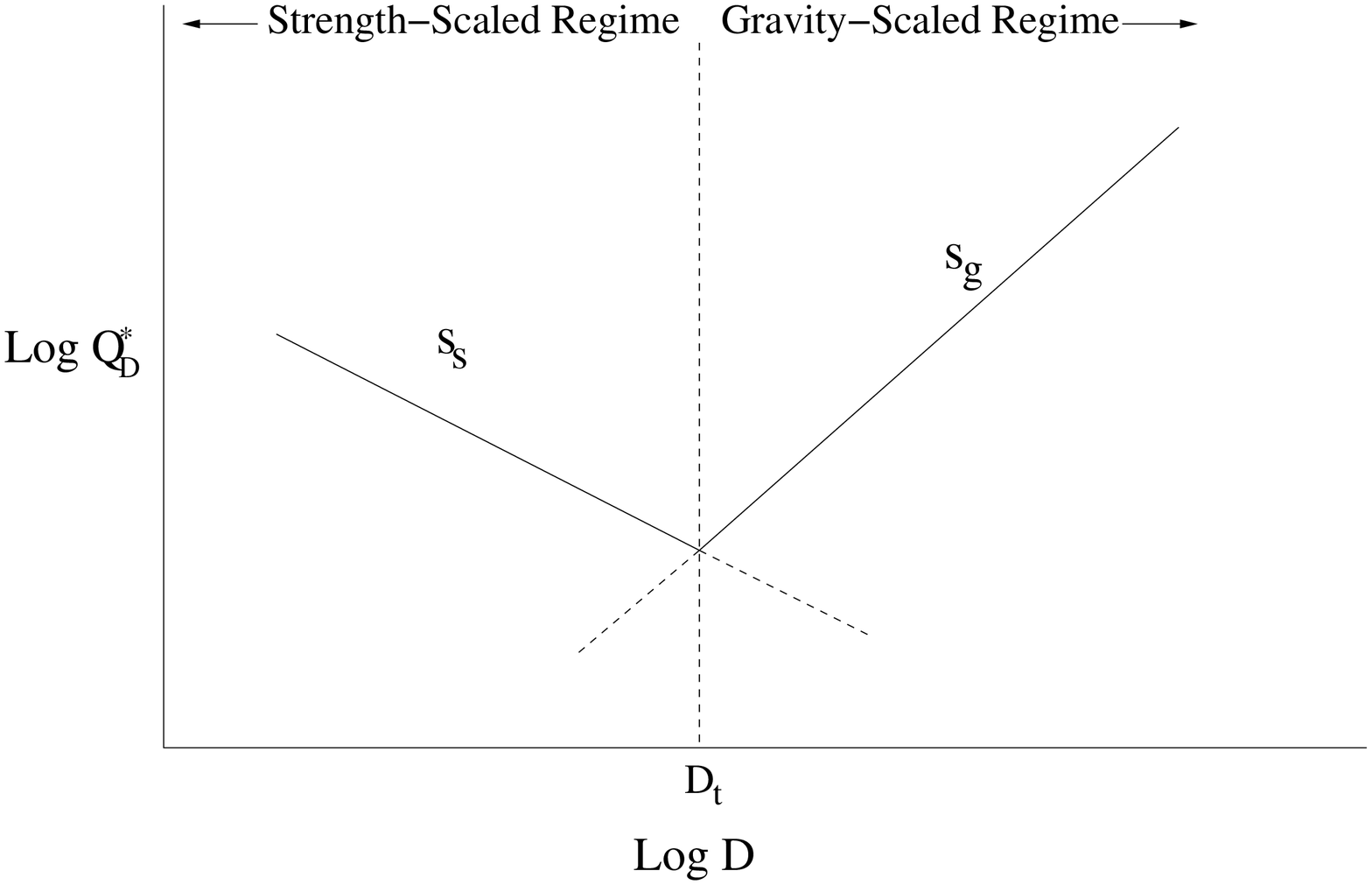}
\centerline{ }
\centerline{ }
\centerline{\Large{Figure 2}}
\end{figure}

\clearpage

\begin{figure}
\plotone{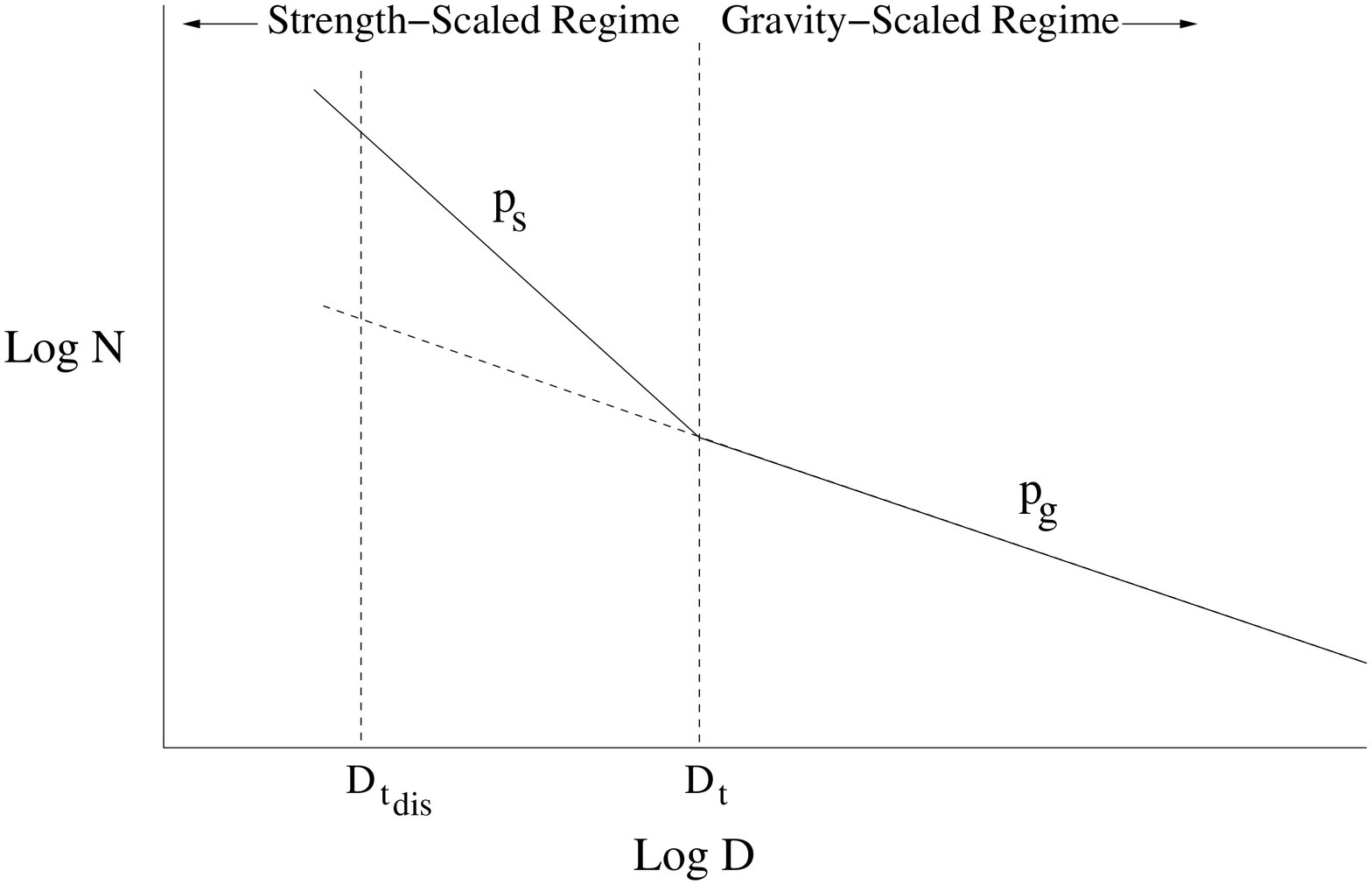}
\centerline{ }
\centerline{ }
\centerline{\Large{Figure 3a}}
\end{figure}

\clearpage

\begin{figure}
\plotone{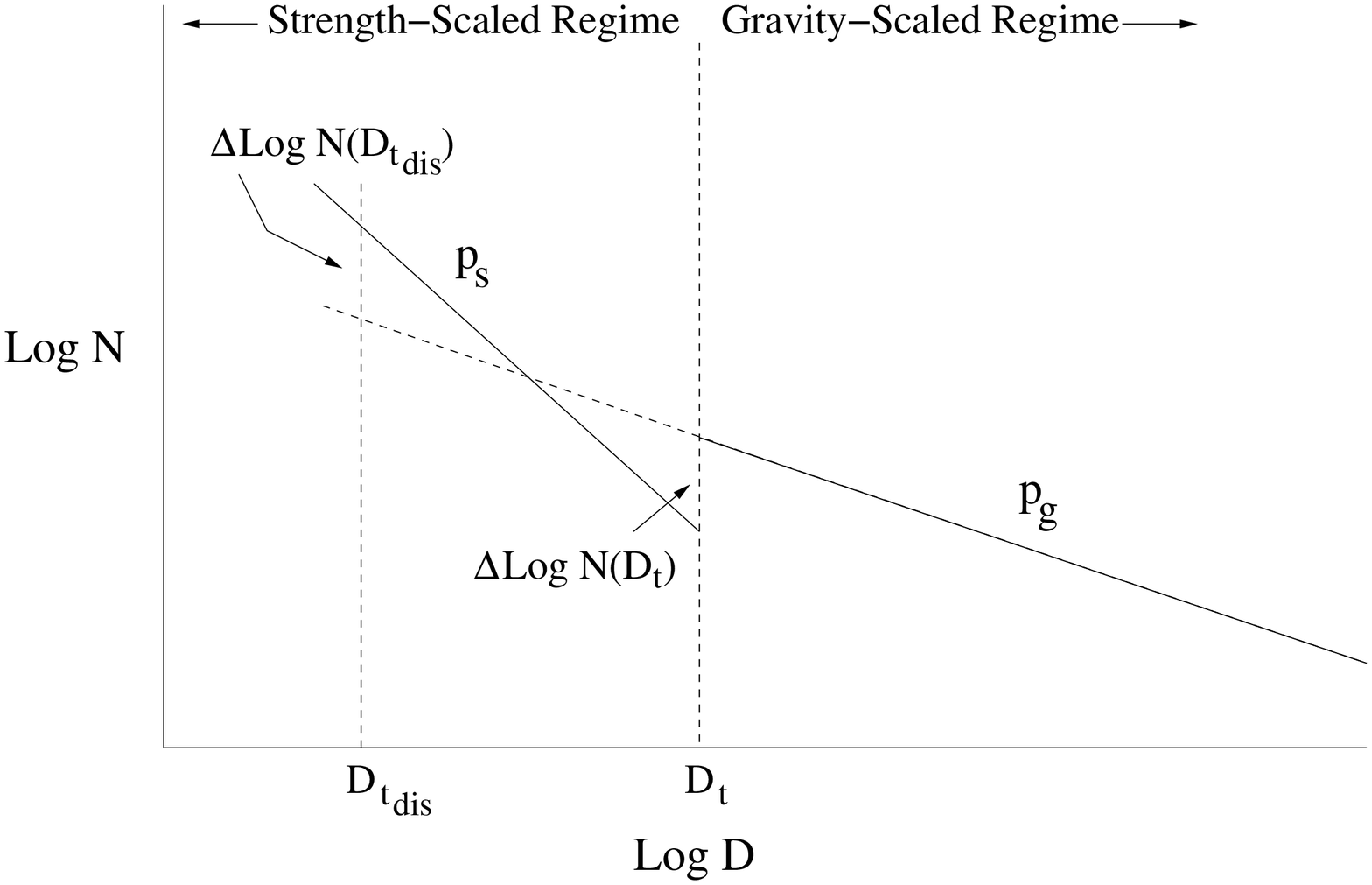}
\centerline{ }
\centerline{ }
\centerline{\Large{Figure 3b}}
\end{figure}

\clearpage

\begin{figure}
\plotone{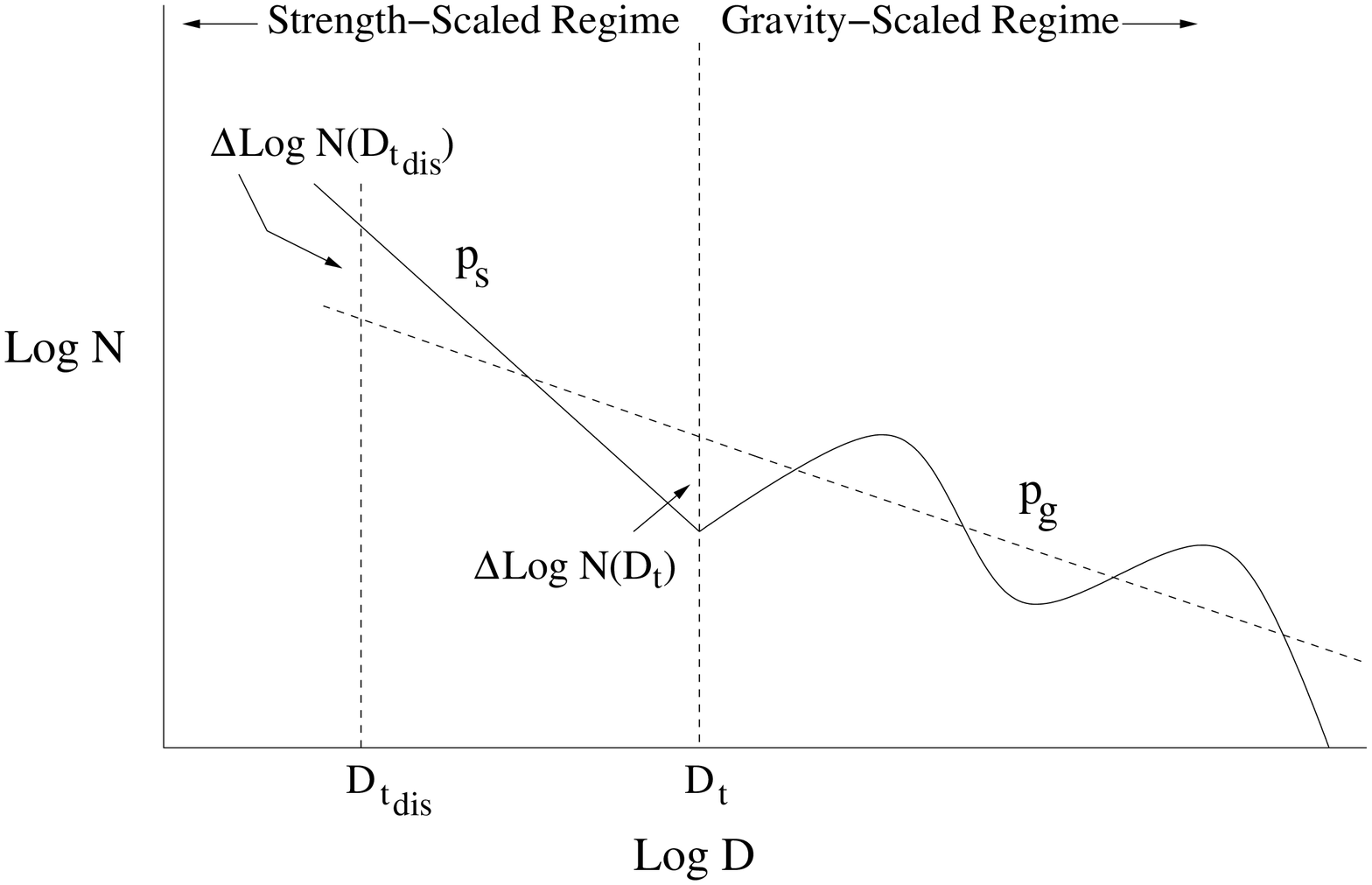}
\centerline{ }
\centerline{ }
\centerline{\Large{Figure 3c}}
\end{figure}

\clearpage

\begin{figure}
\plotone{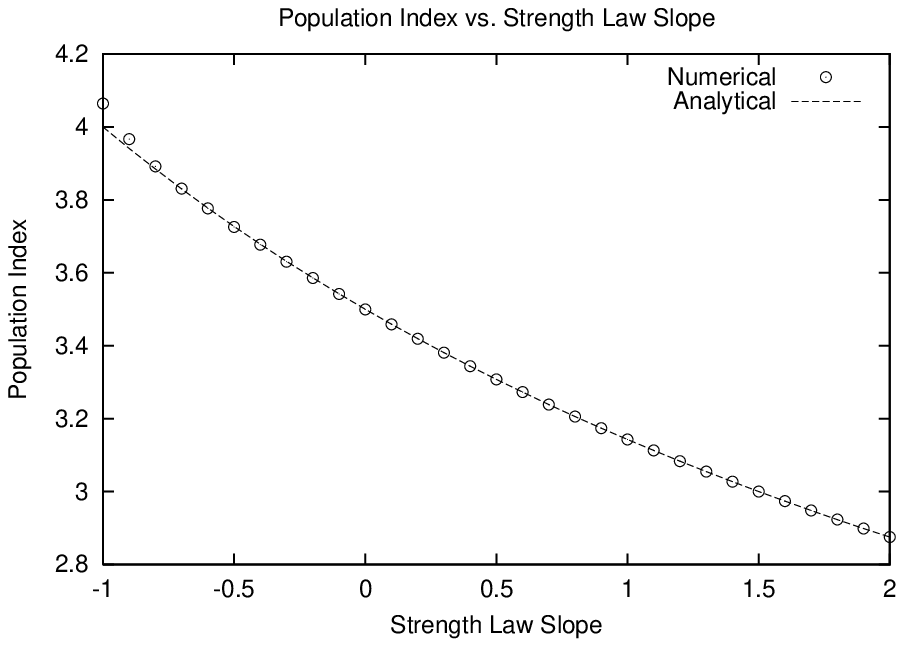}
\centerline{ }
\centerline{ }
\centerline{\Large{Figure 4}}
\end{figure}

\clearpage

\begin{figure}
\plotone{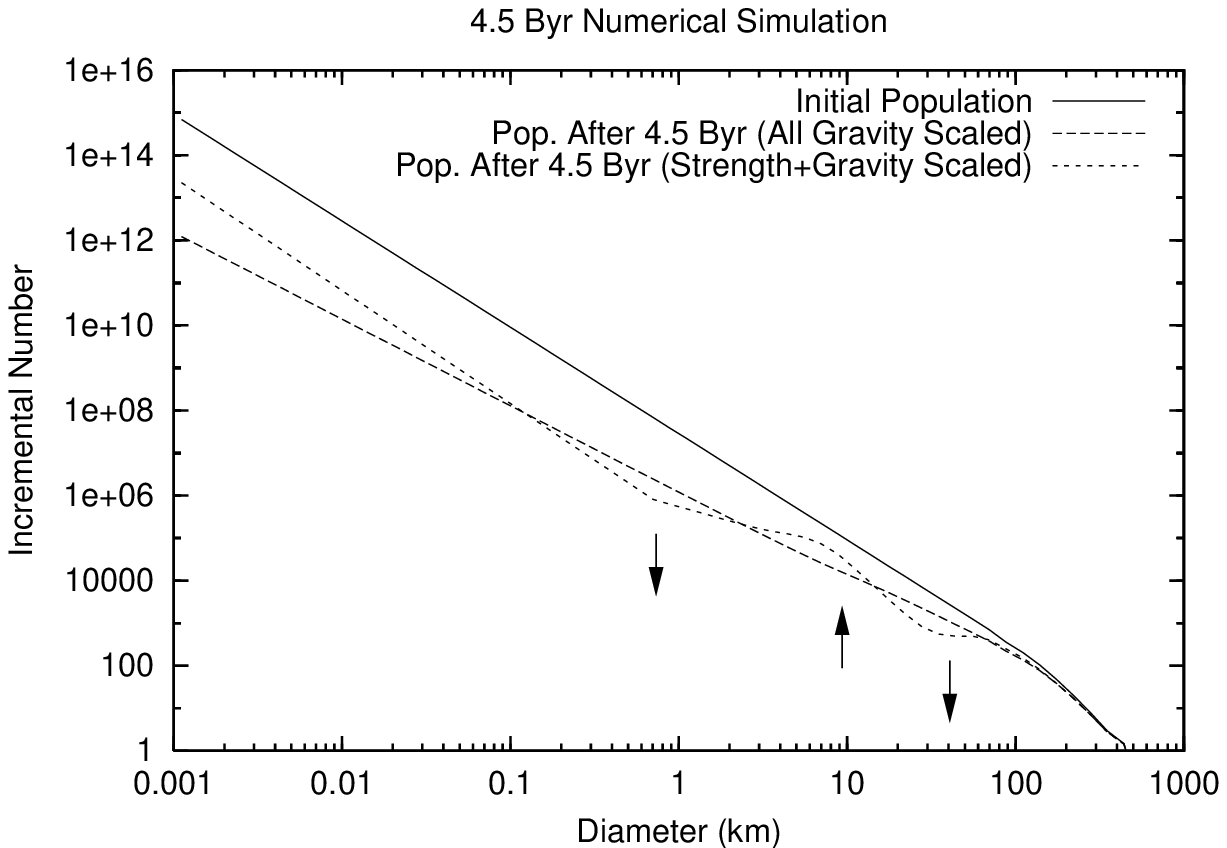}
\centerline{ }
\centerline{ }
\centerline{\Large{Figure 5}}
\end{figure}

\clearpage

\begin{figure}
\plotone{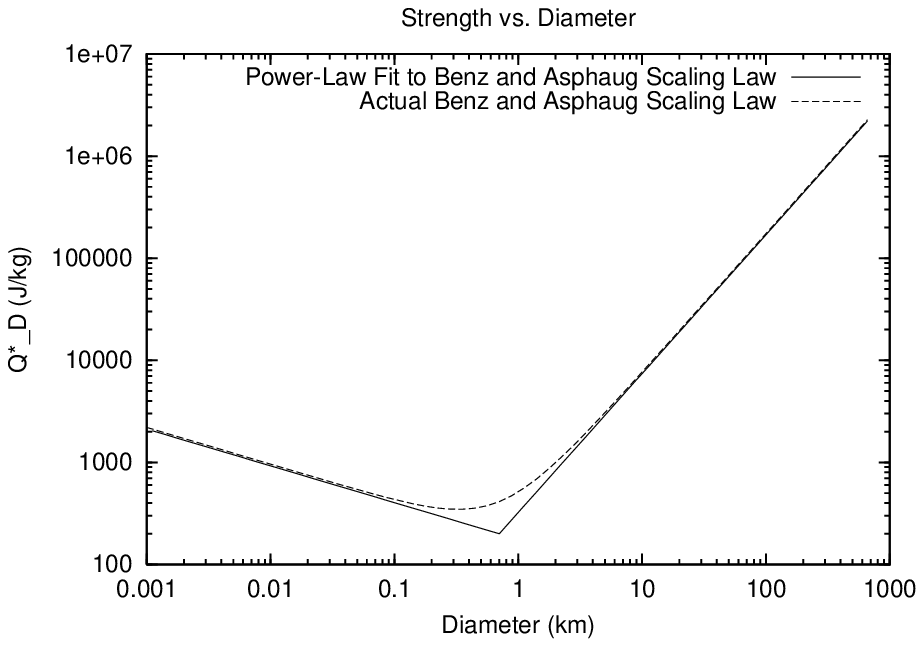}
\centerline{ }
\centerline{ }
\centerline{\Large{Figure 6}}
\end{figure}

\end{document}